\documentclass[Afour,sagev,times]{sagej}

\usepackage{moreverb,url}
\usepackage{float}
\usepackage{amsmath}

\usepackage[colorlinks,bookmarksopen,bookmarksnumbered,citecolor=red,urlcolor=red]{hyperref}

\newcommand\BibTeX{{\rmfamily B\kern-.05em \textsc{i\kern-.025em b}\kern-.08em
T\kern-.1667em\lower.7ex\hbox{E}\kern-.125emX}}

\def\volumeyear{20XX}
\usepackage{xr}
\begin{document}

\runninghead{Chandereng and Chappell}

\title{Robust Blocked Response-Adaptive Randomization Designs}

\author{Thevaa Chandereng\affilnum{1,2} and Rick Chappell\affilnum{1,2}}

\affiliation{\affilnum{1}Department of Biostatistics \& Medical Informatics, University of Wisconsin-Madison, Madison, WI, USA\\
\affilnum{2}Department of Statistics, University of Wisconsin-Madison, Madison, WI, USA}

\corrauth{Thevaa Chandereng, 
Department of Biostatistics \& Medical Informatics,
University of Wisconsin-Madison,
201 WARF, 610 Walnut Street, Madison, WI 53726.}

\email{chandereng@wisc.edu}

\begin{abstract}
In most clinical trials, patients are randomized with equal probability among treatments to obtain an unbiased estimate of the treatment effect. 
Response-adaptive randomization (RAR) has been proposed for ethical reasons, where the randomization ratio is tilted successively to favor the better performing treatment. 
However, the substantial disagreement regarding bias due to time-trends in adaptive randomization is not fully recognized.
The type-I error is inflated in the traditional Bayesian RAR approaches when a time-trend is present. 
In our approach, patients are assigned in blocks and the randomization ratio is recomputed for blocks rather than traditional adaptive randomization where it is done per patient. 
We further investigate the design with a range of scenarios for both frequentist and Bayesian designs. 
We compare our method with equal randomization and with different numbers of blocks including the traditional RAR design where randomization ratio is altered patient by patient basis. 
The analysis is stratified if there are two or more patients in each block. 
Small blocks should be avoided due to the possibility of not acquiring any information from the $\mu_i$. 
On the other hand, RAR with large blocks has a good balance between efficiency and treating more subjects to the better-performing treatment, while retaining blocked RAR's unique unbiasedness.

\end{abstract}


\maketitle

\section{Introduction}

\subsection{Response-adaptive randomization}
Randomization remains a pivotal methodology for advancement in medical knowledge properly done. 
It removes any systematic bias and allows direct inference between the treatment group and outcome. 
Traditionally, a fixed randomization scheme (usually 1:1 or 2:1)  is used to due to simplicity in design and execution of the trial. 
However, response-adaptive randomization (RAR) designs utilize accrual information to adaptively tilt the randomization ratio to the better performing treatment group.
Patients enrolled in these trials are treated to the best way possible, according to current interim estimates \citep{wei1978randomized}.
However, in traditional RAR confounding of treatment with time induces a potentially severe bias \citep{begg2015ethical, thall2015statistical, chappell2007continuous, karrison2003group}. 
The purpose of this article is to expand on the characteristics of blocked RAR, proposed by Karrison et al. as a way to eliminate this bias \citep{karrison2003group}. 
Although, in this paper we focus on trials with two parallel intervention groups, our method are easily extendable to three or more arms.  

RAR has been highly advocated due to ethical reasons. 
It's primary is that it improves the benefit/risk for the patients enrolled in a trial \citep{hey2015outcome}. 
For instance, in Zidovudine Treatment (AZT) trial conducted to test the reduction of maternal-infant transmission of HIV type 1, patients were randomized 1:1 \citep{connor1994reduction}.
The equal randomization scheme placed 239 women in the treatment group (AZT) and 238 patients in the placebo group. 
Out of the 238 women in the AZT group, 60 of the infants were transmitted with the HIV virus while only 20 infants contracted the HIV virus for the AZT group \citep{connor1994reduction}. 
The outcome of the trial confirmed that the new treatment works.
Given the seriousness of the outcome of this study, it is reasonable to argue that 50-50 allocation was unethical. 
As the outcome of the trial becomes available, the randomization ratio should have tilted in favor of the AZT group. 
RAR design could have reduced the number of infants that contracted HIV disease from their maternal. 

On the other hand, opponents of RAR have argued that adaptive randomization challenges the whole notion of equipoise \cite{begg2015ethical}. 
Hey and Kimmelman also argued that most new treatments offer small improvement over standard treatments, thus they offer limited benefit and require a larger sample size \cite{hey2015outcome}.
Hey and Kimmelman also suggested that equal randomization helps reduce the trial size and length, thus it benefits future patients rather than current patients enrolled in the trial \cite{hey2015outcome}. 
Korn and Friedlin measure the difference in non-responders under equal and adaptive randomization and found that adaptive randomization required a larger trial to achieve the same power and type-I error \cite{korn2011outcome}. 
Also, outcomes in RAR trials must be short to be able to obtain the outcome of the trial for future randomization \citep{karrison2003group}.

\subsection{Time-trend issues}

As stated above, a major criticism of RAR is the time-trend issue. 
This is a main factor for why RAR is infrequently used. 
The type-I error rate is usually not controlled at the nominal level under traditional Bayesian or frequentist RAR designs \citep{thall2015statistical}. 
Besides affecting type-I error, studies have shown that there is a large bias in the estimation of treatment difference under traditional RAR designs\citep{thall2015statistical}. 
%Figure~\ref{plot:RARplot} shows that the increase in response in the treatment group tilts the randomization ratio in favor of the treatment. 
%However, this is directly confounded by time. 

%\begin{figure}
%\centering
%\includegraphics[width=0.45\textwidth]{RAR_plot.pdf}
%\caption{Effect of time as a blocking factor on randomization ratio}
%\label{plot:RARplot}
%\end{figure}

Examples of time trend issues in RAR design include \citep{chappell2007continuous, karrison2003group}:

\begin{itemize}
\item The disease itself can change, sometimes radically (e.g., AIDS in the early 1990s).
\item Our definition of the disease can change due to new scientific discoveries or diagnostic methods (e.g., stage migration due to introduction of new diagnostic technologies).
\item Inclusion criteria can change, either formally (in which case we can stratify analysis on before vs. after the change) or informally due to “recruiting zeal” or other issues (in which case we can’t).
\item Participating institutions can change, such as when they even into multi-center trials at different times
\item Patients within centers can change, especially but not only with chronic diseases, due to the phenomenon of a queue of desperate patients lining up at the door.
\item In addition to these examples, an investigator who wants to game the system could cross his/her fingers that his favored treatment arm is ahead, then progressively enroll better prognosis patients over time
\end{itemize}

In long duration trials, time-trends are especially likely to occur. 
Patients' characteristics might be completely different throughout the trial or even at the beginning and end of the trial (which is also known as ``patient drift'') \citep{karrison2003group}. 
However, standard RAR analyses assume that the sequence of patients who arrive for entry into the trial represents samples drawn at random from two homogenous populations, with no drift in the probabilities of success” \citep{begg2015ethical, karrison2003group}.
This assumption is usually violated. 
For example, there were more smokers enrolled in the latter part of the trial than the beginning of the trial in the Lung Cancer Elimination(BATTLE) \citep{liu2015overview}.   
Kalish and Begg (1987) noted that in a sampling of large randomized Eastern Cooperative Oncology Group trials moderate time-trends in overall outcomes are common \citep{kalish1987impact}. 

Despite the serious flaw in RAR design, there is not much literature on this area to address the time-trend issue. 
A randomization test for adjusting type-I error inflation was proposed by Simon and Simon using different RAR rules for double-arm trials \citep{simon2011using}. 
Jennison and Turnbull explored a group-sequential method for RAR with continuous outcome utilizing multi-stage
terminology \cite{jennison1999group}.
Karrison et al. introduced a stratified group-sequential method with a simple example of altering the randomization ratio to address this issue \citep{karrison2003group}. 
Coad used a very similar stratified analysis to Karrison et al. \citep{coad1992comparative}.
Rosenberger et al. introduced a covariate-adjusted RAR procedure where time mechanism can be added as a covariate in the model, at the expense of additional assumptions \citep{rosenberger2001covariate}. 
Thall et al. investigated type-I error under a linear time-trend induced in the traditional RAR design and showed that it's type-I error is significantly above the nominal level \citep{thall2015statistical}. 
Villar et al. explored the hypothesis testing procedure adjusting for covariates for correcting type-I error inflation and the effect on power in RAR design with time-trend effect added to the model for two-armed and multi-armed trials \citep{villar2018response}. 

Time-trend can not only greatly bias the estimated in treatment effect but it can also wrongly reject a true null hypothesis. 
We propose a block (group-sequential) design where the randomization ratio is altered in a block level instead of a patient by patient basis using both frequentist and Bayesian approaches. 
The randomization ratio is kept constant in each block.
The block design is similar to the stratified group design introduced by Karrison et al. \citep{karrison2003group}. 
We further study the robustness in different block sizes using both frequentist and Bayesian approach. 
We also compare these results with traditional RAR design and with fixed (1:1) randomization.

\section{Trial Design and Simulation}
\subsection{Block Design for RAR and Why?}
We assume binary outcomes, in which events (success/failure) are observed within a short period from the beginning of the treatment. 
In a block design, patients are enrolled in a sequential manner.  
For instance, in a two-arm design (treatments A and B) patients are enrolled in block with sample size of $n_{Ak}$ and $n_{Bk}$, for $k = 1, 2, 3, ... , K$, where $n_{Ak}$ and $n_{Bk}$ represent the sample sizes in treatment groups A and B in block $k$.
In this design, the randomization ratio is constant for patients within each block and the randomization ratio is altered at the block level. 
Unlike traditional RAR which potentially alters the randomization ratio on an individual basis, this method speeds up the process of RAR trials since the randomization ratio is modified for a group of patients in the block. 
The initial randomization ratio is set to 1:1.
As asserted by Karrison et al., the block design will eliminate bias due to drift through stratification \citep{karrison2003group}. 
However, the optimal block size remains unclear.
Blocks with a large number of patients also help reduce the probability of imbalance in the wrong direction, where more patients are assigned to the inferior treatment compared to traditional RAR.
Blocks with few patients can reduce the power of the trial because if patients are not randomized to both treatments in a block, the block becomes uninformative. 
 
\subsection{Simulation and Design}
We investigated the effect of various numbers of blocks using simulations for both frequentist and Bayesian approach. 
The rules for altering the randomization ratio in both Bayesian and frequentist designs are described in the subsection below.
The target sample size of  $N$ = 200 subjects was considered with number of blocks (strata), $K$ = 1, 2, 4, 5, 10, 20, 100, 200. 
When $K$ =  1 this follows the traditional equal allocation design and when $K$ = 200 this follows the traditional RAR design without stratification. 
Upon completion of enrollment and collection of data on the subjects in each block, an interim analysis is done to revise the allocation rule. 
The interim analysis also allows for early stopping and details of early stopping for both Bayesian and frequentist designs are included below. 
10, 000 independent simulations were performed for each design yielding a maximum Monte-Carlo standard error of 0.25\%. 

In each simulation, the success rate of treatment A (control group), $p_A$ is set to 0.25. 
The alternative (the success rate of treatment B, $p_B$) is set to 0.25 (null case), 0.35 and 0.45.
On top of simulating both Bayesian and frequentist designs for constant $p_A$ and $p_B$ specified above, we also simulated RAR with drift effect. 
To examine the effects of drift, we increased both $p_A$ and $p_B$ linearly from their initial values to a final value of 0.25 larger. 
The drift was added on a patient by patient basis. 
For instance, the success rate in patient n for treatment A is $p_A(n) = 0.25 + 0.25(n / 200)$, when $n$ patient had been accrued.
While both the $p_A$ and $p_B$, the treatment effect remains constant throughout the trial.
For both frequentist and Bayesian designs, the final analysis is done using stratification for a number of blocks K, where $ 1< K < 200$. 
For traditional RAR and ordinary with 1:1 allocation ratio the standard analyses are used. 
Details of the analysis are attached in the Bayesian and frequentist approach sections.

\subsection{Frequentist Approach}

The allocation rule for the frequentist approach is altered using an optimal allocation ratio for 2-armed trial specified by Rosenberger et al \cite{rosenberger2001optimal}.
The allocation probability for treatment A in stratum $j$ is defined as 
$$\pi_{j, A} = \frac{\sqrt{\hat{p}_{A,j - 1} }}{\sqrt{\hat{p}_{A, j- 1}} + \sqrt{\hat{p}_{B,j - 1}}},$$
where $\hat{p}_{A, j - 1}$ is the estimated success rate of treatment A and $\hat{p}_{B, j - 1}$ is the estimated success rate of treatment B after block $j - 1$. 
Simulation for early stopping is also included. 
The alpha spending approach was incorporated to stop for early success or failure \cite{demets1994interim}. 
For traditional RAR and fixed allocation, a one-sided chi-square test was used to analyze the outcome. 
For the block design, a one-sided Cochran-Mantel-Haenszel test was utilized to deal with the stratification. 
Yates's correction was not applied for both chi-square and Cochran-Mantel-Haenszel test due to the overly conservative nature of the correction \cite{haviland1990yates}. 
Treatment B is proclaimed superior to treatment A at the trial's termination if the one-sided p-value $< 0.05$.

The estimated treatment difference is computed as below for stratified design. 
In the two treatment scenario ( A \& B), suppose there are $K$ strata, let $p_{Ak}$ and $p_{Bk}$ be the proportions of success and $n_{Ak}$ and $n_{Bk}$ the numbers of patients in stratum $k$. 
Let
$$\hat{\delta_k} = \hat{p_{Bk}}- \hat{p_{Ak}}$$
be the observed difference in proportion between treatment A and B in stratum $k$. 
The estimated treatment difference ($\delta$) is  
$$\hat{\delta} = \sum^K_{k = 1} w_k \hat{\delta_k}$$
where $w_k = \frac{(n_{Ak}^{- 1} + n_{Bk}^{- 1})^ {-1}}{\sum_{k = 1}^ K(n_{Ak}^{- 1} + n_{Bk}^{- 1})^ {-1}} $ is the weight of stratum k and $\sum_{k = 1}^ K w_k = 1$. 
For a non-stratified design, the corresponding estimates is
$$\hat{\delta} = \hat{p_B} -  \hat{p_A}$$, 
where $p_A$ and $p_B$ are the sample proportion of success in treatment A and B.

\subsection{Bayesian Approach}

In the Bayesian approach, the Bayesian adaptive randomization (BAR(1/2)) method used by Thall and Wathen is employed \cite{thall2007practical}. 
The probability of randomizing subjects to treatment A in stratum j, $\pi_{j, A}$ is defined as
$$\pi_{j, A} = \frac{(p_{A>B, j - 1} (data))^ {1/2}}{(p_{A>B, j - 1} (data))^ {1/2} + (p_{B>A, j - 1} (data))^ {1/2}},$$
where $p_{A > B, j - 1}(data)$ is the posterior probability that treatment A has a higher success rate than  treatment B and $p_{A>B, j - 1} (data)) = 1 - p_{B>A, j - 1} (data))$ after the $j - 1 ^{st}$ block . 
The beta-binomial conjugate prior is used for the estimation of the posterior probability. 
The posterior probability that the treatment A has a higher event rate than treatment B is
\begin{equation*}
\begin{split}
&  p_{A > B, j - 1}(data) \sim  \\
& beta(y_{A, j - 1} + a_0, N_{A, j - 1}  - y_{A, j - 1}  + b_0) -  \\
& beta(y_{B, j - 1}  + a_0,  N_{B, j - 1} - y_{B, j - 1} + b_0) > 0,
\end{split}
\end{equation*} 
where $y_{A, j - 1}$ and $y_{B, j - 1}$ denote the numbers of events in treatment A and B, $N_{A, j - 1}$ and $N_{B, j - 1}$ denote the numbers of subjects in treatment A and B after block j - 1, $a_0$ and $b_0$ denote the prior rate parameter of the beta distribution. 
Similar to Thall and Wathen's approach, $Beta(0.5, 0.5)$ priors were assumed for both treatments A and B \cite{thall2007practical}. 
Since there is no closed-form solution for the difference in beta distributed random variables Monte Carlo simulations were performed to estimate the posterior of the treatment difference.

Similar to the frequentist design, simulations for the possibility of early stopping are included.  
If $P_{B > A, j - 1}(data) > .99$ the trial is stopped early for success and if $P_{B>A, j - 1}(data) < .01$ the trial is stopped early for failure. 
For non-stratified design, if the final posterior probability $P_{B > A}(data) > .95$  treatment B is declared superior to treatment A. 
The mean treatment difference is estimated using Monte Carlo simulation using the mean of $P_{B>A}(data)$. 

For block-stratified design, Bayesian logistic regression was implemented to estimate the posterior probability of treatment difference. 
The logistic regression model is defined as below
$$p \sim \beta_0 + \beta_{trt} x_{trt} +  \beta_{1}x_1 + ..+ \beta_{K}x_k , $$
where $p$ probability of success, $\beta_0$ is the intercept term, $\beta_{trt}$ is the treatment effect, $x_{trt}$ is the indicator variable for treatment, $\beta_j$ is the stratum effect of stratum $j$, and $x_j$ is the indicator variable of stratum $j$, for j = 1,.., K. 
The uninformative prior applied in the model are as defined
$$\beta_0, \beta_{trt}, \beta_j \sim N(0, \sigma^2_j), \qquad \sigma_j^2 \sim Inv-\chi^2(1, 2.5).$$
The posterior value of $\beta_{trt}$ is used to estimate the proportion difference in treatment. 
The R package \textit{arm} with \textit{quasi} family and identity link with $\mu(1-\mu)$ variance was used to fit Bayesian logistic regression and obtain the posterior samples of $\beta_{trt}$ \cite{arm}.
The treatment effect is estimated using the mean posterior value of $\beta_{trt}$. 
Treatment B is declared superior to treatment A if $P(\beta_{trt} > 0) > 0.95$ since $\beta_{trt}$ computes the estimated difference in proportion between treatment B and treatment A .  

\section{Results}
\begin{figure}
\centering
\includegraphics[width=0.5\textwidth]{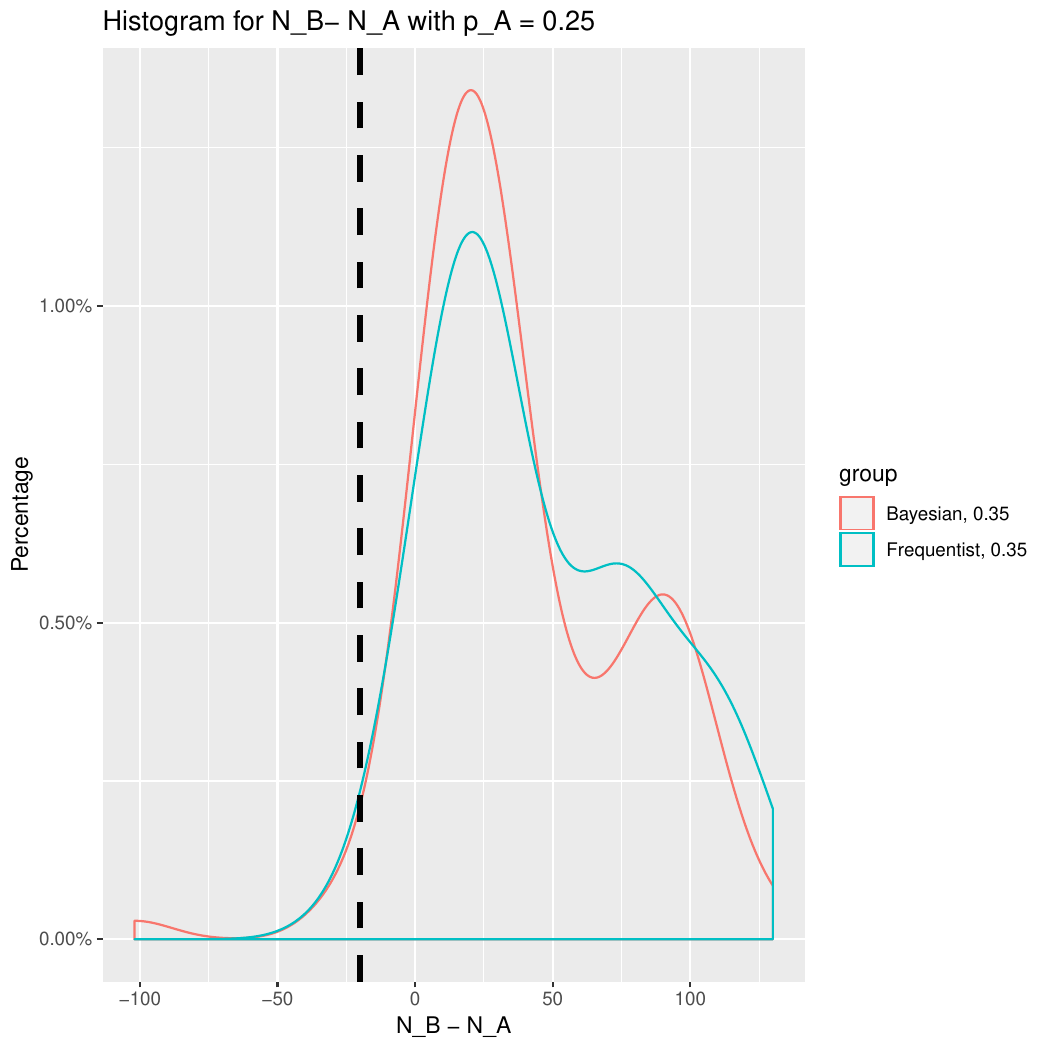}
\caption{Distribution of difference in number of patients assigned to treatment B compared to treatment A ($N_B- N_A$) for a 200 patient trial with no early stopping using both Bayesian and frequentist stratified designs with 10 number of blocks. The  sample size difference is collected for 100 different simulation with $p_A = 0.25$ and $p_B = 0.35$. The black dashed line included shows $\pi_{20}$. }
\label{plot:ssdiff}
\end{figure}

For all cases, the simulation was replicated 10,000 times for both Bayesian and frequentist designs with and without time trends. 
The number of blocks are K = 1, 2, 4, 5, 10, 20, 100, 200.  
For all the simulations, the power (probability of correctly concluding that treatment B is superior to treatment A), bias, probability of sample imbalance of more than 20 patients assigned to treatment A over treatment B, the mean, 2.5\%, and 97.5\% percentile of difference between $N_B - N_A$ (sample size imbalance favoring treatment B over treatment A) are reported.
The power reflects the proportion of 10,000 trials that declares treatment B superior to treatment A. 
The type-I error (false-positive) is simulated by setting both  $p_A, p_B = 0.25$.
The bias $\hat{\delta} - \delta$.
$\pi_{20}$, the probability of assigning 20 or more subjects to treatment A over treatment B ($P(N_A -N_B > 20)$) is included in the simulation. 
Due to the nontrivial possibility of assigning more patients to the inferior arm by chance, it is vital to analyze $\pi_{20}$. 
To highlight the main advantage of RAR compared to equal randomization, the difference in sample size between the superior arm and inferior arm is presented.
Since the difference between subjects assigned to treatment B and treatment A ($N_B - N_A$) is skewed and dispersed as illustrated in Figure~\ref{plot:ssdiff}, the mean, 2.5\% and 97.5\% of $N_B- N_A$ are reported.

\subsection{Simulation with No Time-Trend}

Tables~\ref{freqnoearlystopnodrift} \&~\ref{bayesnoearlystopnodrift} displays the results of simulations for numbers of blocks, K = 1, 2, 4, 5, 10, 20, 100, 200 using the frequentist and Bayesian approaches with no drift applied and not stopping early for success or failure.
The frequentist approach (Table~\ref{freqnoearlystopnodrift}) manages to control the type-I error unlike the Bayesian design in Table~\ref{bayesnoearlystopnodrift}. 
Even though the stratified RAR design has higher type-I error, the type-I error is still controlled under the nominal level (Table~\ref{freqnoearlystopnodrift}).
However, the false-positive rate is high for 4, 5, 10 and 20 blocks under the Bayesian design (Table~\ref{bayesnoearlystopnodrift}). 
Block design with 2, 4, 5 and 10 number of blocks provides a higher power when $p_B = 0.35, 0.45$ (Table~\ref{freqnoearlystopnodrift}). 
In the Bayesian design, the fixed randomization ratio provides the best power. 
Although the type-I error is close to the nominal level under K = 2, 4 and 5, with a slight increase in sample size it could be lowered to the nominal level. 
A small increase in sample size might still be favorable if more subjects are treated with the best possible cure. 
Blocked designs with a small number of subjects in a block should not be considered due to low power as shown in Tables~\ref{freqnoearlystopnodrift} and \ref{bayesnoearlystopnodrift} with 100 blocks. 
This poor performance is due to the reality that the blocks will be noninformative if all subjects in the block receive the same treatment. 
Ethically, this design (2 subjects per block) places more subjects at risk without contributing to the advancement in science.

\begin{table}[ht]
\resizebox{\columnwidth}{!}{%
\begin{tabular}{|c|c|c|c|c|c|c|}
\hline
$p_B$ & Block & Size/Power & Bias & $\pi_{20}$ & $N_B - N_A$ \\
\hline
0.25 & 1 & 0.03 & 0.00 & 0.07 & -0.35 (-28, 28) \\ 
   & 2 & 0.05 & 0.00 & 0.10 & 0.22 (-34, 34) \\ 
   & 4 & 0.05 & 0.00 & 0.14 & 0.03 (-40, 40) \\ 
   & 5 & 0.05 & 0.00 & 0.15 & -0.06 (-40, 40) \\ 
   & 10 & 0.05 & 0.00 & 0.16 & 0.08 (-42, 42) \\ 
   & 20 & 0.06 & 0.00 & 0.16 & 0.03 (-42, 42) \\ 
   & 100 & 0.05 & 0.00 & 0.17 & -0.25 (-44, 42) \\ 
   & 200 & 0.02 & 0.00 & 0.16 & 0.12 (-42, 44) \\ 
   \hline
 0.35 & 1 & 0.34 & 0.00 & 0.07 & 0.27 (-28,  28) \\ 
   & 2 & 0.46 & 0.00 & 0.04 & 8.64 (-24, 40) \\ 
   & 4 & 0.46 & 0.00 & 0.03 & 13.26 (-24, 52) \\ 
   & 5 & 0.44 & 0.00 & 0.03 & 13.85 (-24, 52) \\ 
   & 10 & 0.44 & 0.00 & 0.04 & 15.24 (-24, 56) \\ 
   & 20 & 0.43 & 0.00 & 0.03 & 16.04 (-24, 56) \\ 
   & 100 & 0.20 & 0.00 & 0.03 & 16.39 (-24, 56) \\ 
   & 200 & 0.34 & 0.00 & 0.03 & 16.32 (-24, 58) \\ 
   \hline
 0.45 & 1 & 0.85 & 0.00 & 0.07 & 0.19 (-28, 28) \\ 
   & 2 & 0.91 & 0.00 & 0.01 & 14.64 (-16, 46) \\ 
   & 4 & 0.91 & 0.00 & 0.01 & 23.13 (-12, 60) \\ 
   & 5 & 0.90 & 0.00 & 0.01 & 24.80 (-12, 62) \\ 
   & 10 & 0.89 & 0.00 & 0.01 & 27.24 (-10,  66) \\ 
   & 20 & 0.88 & 0.00 & 0.00 & 27.90 (-10, 66) \\ 
   & 100 & 0.49 & 0.00 & 0.00 & 28.79 (-10, 68) \\ 
   & 200 & 0.84 & 0.00 & 0.01 & 28.85 (-10, 70) \\ 
\hline
\end{tabular}
}
\caption{RAR using frequentist approach with no early stopping criteria and no drift. $p_A$ is set to 0.25 for all cases. $Bias = \hat{\delta} - \delta$. $\pi_{20}$ denote the probability of assigning more than 20 patients in the inferior treatment group. $N_A$ and $N_B$ denotes the number of patients assigned to treatment A and B. The mean (2.5\%, 97.5\%) of $N_B - N_A$ is reported in the last column. 10,000 simulations were done for each case.}
\label{freqnoearlystopnodrift}
\end{table}

The estimated treatment effect is unbiased under the frequentist design regardless of the number of blocks used.
However, it is biased for most stratified RAR designs and traditional RAR as shown in Table~\ref{bayesnoearlystopnodrift}. 
Variability in sample size assigned to treatment A and B is smaller in the frequentist approach (Table~\ref{freqnoearlystopnodrift}) than the Bayesian approach (Table~\ref{bayesnoearlystopnodrift}). 
The mean difference in the subject's assignment of treatment is also smaller in the frequentist design compared to the Bayesian design. 
Thus, the frequentist design is more conservative in assigning patients to the better-performing treatment than the Bayesian design. 
At the largest difference in proportion $p_B = 0.45$, $p_A = 0.25$, there is a possibility of imbalance in the wrong direction in both designs. 
The probability of imbalance in sample size favoring the inferior treatment is small under all scenarios as shown by $\pi_{20}$ in Table~\ref{freqnoearlystopnodrift}.
The difference in sample size ($N_B - N_A$) is relatively small and close to 0 for the frequentist design. 
On the other hand, the imbalance is large for the Bayesian design as illustrated by $\pi_{20}$ in Table~\ref{bayesnoearlystopnodrift}.
The large difference in sample size (more than half the total) is seen in the Bayesian design. 

Tables~\ref{freqearlystopwnodrift}  \&~\ref{bayesearlystopnodrift} display the results of simulations for K = 1, 2, 4, 5, 10, 20, 100, 200 using frequentist and Bayesian approaches and with early stopping criteria for success or failure implemented.
Parallel to the earlier results, the bias is higher in Tables~\ref{freqearlystopwnodrift} and ~\ref{bayesearlystopnodrift}.

%Table~\ref{bayesearlystopwdrift} emphasizes that K = 10, 20, 100 and 200 have a inflated type-I error of 0.12, 0.16 and 0.18  similar to Table~\ref{bayesnoearlystopwdrift}. 
%As seen in type-I error and power in Table~\ref{freqearlystopwdrift} and \ref{bayesearlystopwdrift}, a large number of strata with Bayesian design should not be considered for clinical studies.

\begin{table}[ht]
\resizebox{\columnwidth}{!}{%
\begin{tabular}{|c|c|c|c|c|c|}
\hline
$p_B$ & Block & Size/Power & Bias & $\pi_{20}$ &  $N_B - N_A$ \\
\hline
0.25 & 1 & 0.05 & 0.00  & 0.07 & -0.08 (-26, 28) \\ 
   & 2 & 0.06 & 0.00 & 0.32 & -0.54 (-76,  76) \\ 
   & 4 & 0.08 & 0.00 & 0.36 & -1.15 (-108, 104) \\ 
   & 5 & 0.08 & 0.00 & 0.36 & -0.47 (-112, 112) \\ 
   & 10 & 0.13 & 0.01 & 0.37 & 1.39 (-124, 124) \\ 
   & 20 & 0.15 & 0.01 & 0.38 & 0.00 (-130, 130) \\ 
   & 100 & 0.01 & 0.00 & 0.39 & -0.96 (-130, 132) \\ 
   & 200 & 0.07 & 0.00 & 0.39 & -0.10 (-134, 132) \\ 
   \hline
 0.35  & 1 & 0.47 & 0.00 & 0.07 & -0.03 (-28,  28) \\ 
   & 2 & 0.46 & 0.01 & 0.06 & 39.91 (-38,  100) \\ 
   & 4 & 0.45 & 0.01 & 0.07 & 57.26 (-50, 136) \\ 
   & 5 & 0.47 & 0.02 & 0.07 & 61.33 (-52, 144) \\ 
   & 10 & 0.51 & 0.02 & 0.07 & 67.24 (-56, 156) \\ 
   & 20 & 0.58 & 0.01 & 0.07 & 70.34 (-58, 162) \\ 
   & 100 & 0.08 & -0.08 & 0.07 & 72.26 (-62, 164) \\ 
   & 200 & 0.45 & 0.01 & 0.07 & 71.94 (-60, 164) \\ 
   \hline
 0.45  & 1 & 0.91 & 0.00 & 0.07 & -0.04 (-28,  28) \\ 
   & 2& 0.89 & 0.01 & 0.01 & 69.83 (2, 110) \\ 
   & 4 & 0.87 & 0.02 & 0.01 & 99.05 (12, 150) \\ 
   & 5 & 0.87 & 0.03 & 0.01 & 104.18 (16, 156) \\ 
   & 10 & 0.85 & 0.03 & 0.01 & 113.65 (24, 172) \\ 
   & 20 & 0.87 & 0.02 & 0.01 & 117.41 (20, 176) \\ 
   & 100 & 0.29 & -0.12 & 0.01 & 119.57 (26, 176) \\ 
   & 200 & 0.87 & 0.02 & 0.01 & 120.42 (28, 176) \\ 
\hline
\end{tabular}
}
\caption{RAR using Bayesian approach with no early stopping criteria and with no drift. $p_A$ is set to 0.25 for all cases. $E(N)$ represents the mean sample size for 10,000 simulation. $Bias = \hat{\delta} - \delta$. $\pi_{20}$ denote the probability of assigning more than 20 patients in the inferior treatment group. $N_A$ and $N_B$ denotes the number of patients assigned to treatment A and B. The mean (2.5\%, 97.5\%) of $N_B - N_A$ is reported in the last column. 10,000 simulations were done for each case.} 
\label{bayesnoearlystopnodrift}
\end{table}

\subsection{Simulation with Time-Trend}

Tables~\ref{freqnoearlystopwdrift} \&~\ref{bayesnoearlystopwdrift} display the results of simulations for K = 1, 2, 4, 5, 10, 20, 100, 200 using the frequentist and Bayesian approaches with 0.25 drift applied and not stopping early for success or failure.
With time drift, the false positive rate is still under the nominal level for the frequentist design with all block sizes as seen in Table~\ref{freqnoearlystopwdrift}. 
However, the type-I error is inflated for traditional RAR and small blocks in the Bayesian design.
Having 2, 4, 5, 10 and 20 blocks remains the most powerful design with the most extreme difference in outcome of $p_A$ and $p_B$.
In the Bayesian design, block size with 50 subjects is still comparable to the traditional fixed randomization ratio design, since the type-I error is controlled under 0.05 and it has the highest power of 0.91 under the maximum difference between $p_A$ and $p_B$. 
A large number of blocks remains a poor design as illustrated earlier in Table~\ref{bayesnoearlystopnodrift}.
The estimated treatment difference remains similar to the simulation with no time-trend except the bias is a little higher under the traditional RAR design as presented in Table~\ref{bayesnoearlystopwdrift}.
The difference in sample size, the imbalance in the wrong direction and the variability between the arms ($N_B - N_A$) remain comparable to the simulation without time drift applied.

Tables~\ref{freqearlystopwdrift} \&~\ref{bayesearlystopwdrift} displays the results of simulation for K = 1, 2, 4, 5, 10, 20, 100, 200 using the frequentist and Bayesian approach and with early stopping criteria for success or failure implemented.
The output in Table~\ref{freqearlystopwdrift} is similar to that of  Table~\ref{freqnoearlystopwdrift} except the treatment different is slightly biased.
However, it is shown that clinical trials with early stopping are slightly biased compared to trials without early stopping criteria  imposed. 
Table~\ref{bayesearlystopwdrift} emphasizes that K = 10, 20, 100 and 200 have inflated type-I errors of 0.14, 0.17 and 0.11  similar to Table~\ref{bayesnoearlystopwdrift}. 
As seen regarding type-I error and power in Tables~\ref{bayesnoearlystopwdrift} and \ref{bayesearlystopwdrift}, small strata should not be considered.

\begin{table}[ht]
\resizebox{\columnwidth}{!}{%
\begin{tabular}{|c|c|c|c|c|c|}
\hline
$p_B$ & Block & Size/Power & Bias & $\pi_{20}$ & $N_B - N_A$ \\
\hline
0.25 & 1 & 0.02 & 0.00 & 0.07 & -0.08 (-28, 28) \\ 
   & 2 & 0.05 & 0.00 & 0.04 & 0.31 (-32, 32)\\ 
   & 4 & 0.05 & 0.00 & 0.02 & -0.04 (-38, 36) \\ 
   & 5 & 0.05 & 0.00 & 0.02 & -0.04 (-38, 38) \\ 
   & 10 & 0.05 & 0.00 & 0.01 &-0.21 (-38, 38) \\ 
   & 20 & 0.05 & 0.00 & 0.01 & 0.08 (-40, 38) \\ 
   & 100 & 0.05 & 0.00 & 0.01 & -0.02 (-40, 40) \\ 
   & 200 & 0.03 & 0.00 & 0.01 & -0.10 (-40, 38) \\ 
      \hline
0.35  & 1 & 0.30 & 0.00 & 0.07 & -0.27 (-28, 28) \\ 
   & 2 & 0.41 & 0.00 & 0.04 & 6.85 (-24, 38) \\ 
   & 4 & 0.42 & 0.00 & 0.04 & 11.08 (-24, 48) \\ 
   & 5 & 0.41 & 0.00 & 0.04 & 11.76 (-26, 48) \\ 
   & 10 & 0.40 & 0.00 & 0.04 & 13.18 (-24, 52) \\ 
   & 20 & 0.39 & 0.00 & 0.03 & 13.59 (-24, 52) \\ 
   & 100 & 0.19 & 0.00 & 0.03 & 13.61 (-24, 52) \\ 
   & 200 & 0.30 & 0.00 & 0.04 & 13.37 (-24, 52) \\ 
      \hline
 0.45 & 1 & 0.81 & 0.00 & 0.07 & 0.01 (-28, 28) \\ 
   & 2 & 0.89 & 0.00 & 0.02 & 12.58 (-18, 44) \\ 
   & 4 & 0.89 & 0.00 & 0.01 & 19.35 (-14, 54) \\ 
   & 5 & 0.88 & 0.00 & 0.01 & 21.09 (-12, 56) \\ 
   & 10 & 0.88 & 0.00 & 0.01 & 23.49 (-12, 60) \\ 
   & 20 & 0.86 & 0.00 & 0.01 & 23.90 (-12, 62) \\ 
   & 100 & 0.46 & 0.00 & 0.01 & 24.17 (-12, 60) \\ 
   & 200 & 0.81 & 0.00 & 0.01 & 24.26 (-12, 62) \\ 
\hline
\end{tabular}
}
\caption{RAR using frequentist approach with no early stopping criteria and with 0.25 drift. $p_A$ is set to 0.25 for all cases. $Bias = \hat{\delta} - \delta$. $\pi_{20}$ denote the probability of assigning more than 20 patients in the inferior treatment group. $N_A$ and $N_B$ denotes the number of patients assigned to treatment A and B. The mean (2.5\%, 97.5\%) of $N_B - N_A$ is reported in the last column. 10,000 simulations were done for each case.}
\label{freqnoearlystopwdrift}
\end{table}

\begin{table}[ht]
\resizebox{\columnwidth}{!}{%
\begin{tabular}{|c|c|c|c|c|c|c|}
\hline
$p_B$ & Block & Size/Power & Bias & $\pi_{20}$ & $N_B - N_A$ \\
\hline
0.25& 1 & 0.05 & 0.00 & 0.07 & 0.44 (-28, 28) \\ 
   & 2 & 0.06 & 0.00 & 0.31 & 0.50 (-78, 76) \\ 
   & 4 & 0.08 & 0.00 & 0.37 & 0.41 (-114, 114) \\ 
   & 5 & 0.08 & 0.00 & 0.38 & -0.44  (-120, 122) \\ 
   & 10 & 0.10 & 0.00 & 0.39 & -0.21  (-132, 132) \\ 
   & 20 & 0.14 & 0.01 & 0.39 & -0.53 (-136, 138) \\ 
   & 100 & 0.02 & 0.00 & 0.39 & 0.19 (-140, 140) \\ 
   & 200 & 0.14 & 0.02 & 0.39 & 1.03 (-140, 136) \\ 
  \hline
 0.35 & 1 & 0.41 & 0.00 & 0.07 & -0.20 (-28, 28) \\ 
   & 2 & 0.42 & 0.01 & 0.07 & 38.53 (-40, 100) \\ 
   & 4 & 0.42 & 0.02 & 0.07 & 59.62 (-54, 140) \\ 
   & 5 & 0.42 & 0.02 & 0.08 & 63.88 (-60, 148) \\ 
   & 10 & 0.42 & 0.02 & 0.08 & 69.54 (-64, 162) \\ 
   & 20 & 0.49 & 0.02 & 0.07 & 73.00 (-62, 166) \\ 
   & 100 & 0.11 & -0.07 & 0.08 & 74.59 (-66, 168) \\ 
   & 200 & 0.55 & 0.04 & 0.07 & 75.26 (-62, 168) \\ 
   \hline
 0.45 & 1 & 0.89 & 0.00 & 0.07 & -0.10 (-28, 28) \\ 
   & 2 & 0.86 & 0.01 & 0.01 & 67.82 (-2, 108) \\ 
   & 4 & 0.83 & 0.03 & 0.01 & 101.98 (12, 152) \\ 
   & 5 & 0.83 & 0.03 & 0.01 & 108.53 (16, 160) \\ 
   & 10 & 0.81 & 0.04 & 0.01 & 118.47 (22, 174) \\ 
   & 20 & 0.82 & 0.04 & 0.01 & 122.37 (26, 178) \\ 
   & 100 & 0.32 & -0.10 & 0.01 & 124.20 (30, 180) \\ 
   & 200 & 0.90 & 0.07 & 0.01 & 124.62 (28, 180) \\ 
\hline
\end{tabular}
}
\caption{RAR using Bayesian approach with no early stopping criteria and with 0.25 drift. $p_A$ is set to 0.25 for all cases. $E(N)$ represents the mean sample size for 10,000 simulation. $Bias = \hat{\delta} - \delta$. $\pi_{20}$ denote the probability of assigning more than 20 patients in the inferior treatment group. $N_A$ and $N_B$ denotes the number of patients assigned to treatment A and B. The mean (2.5\%, 97.5\%) of $N_B - N_A$ is reported in the last column. 10,000 simulations were done for each case.}
\label{bayesnoearlystopwdrift}
\end{table}

\section{Conclusion}
RAR designs have the arguably appealing property of assigning more patients to better-performing treatment.
However, trialists need to be careful with issues of time-trend.
Time-trends can significantly impact the type-I error rate which can affect the validity of clinical studies.  
As statisticians, we cannot emphasize the importance of controlling the false-positive rate in a clinical setting.
Thall et al. have shown that methods which able to control type-I error have failed to detect the true treatment difference \citep{thall2015statistical}.

Besides controlling the false positive rate, trialist needs to make sure the method of altering the randomization ratio is not too extreme, affecting the bias and negating the use of randomization in clinical studies. 
Thall et al. have emphasized the difference in simulation between BAR(1) and BAR(1/2), with BAR(1) having a large imbalance in the wrong direction and larger false-positive rate.
Zelen's play-the-winner rule was implemented to an extracorporeal membrane oxygenation (ECMO) trial where the first patient was assigned to both control and treatment group \citep{zelen1969play, mike1993neonatal}.
Due to the failure in the single control and success in the treatment group, all subsequent patients were randomized to the treatment group \citep{mike1993neonatal}.
However, it was later discovered that the first patient assigned to the control group was much sicker than all the patients randomized to the treatment groups. 

On the other hand, scientists need to be aware of the risk that RAR poses which includes having assigned more patients to the inferior treatment. 
As highlight by Thall et al., ``\textit{The practical and ethical point is that AR may behave pathologically in that it carries a nontrivial risk of creating a large sample size imbalance in favor of the inferior treatment}'' \citep{thall2015statistical}.
Large imbalance in the wrong direction can also be controlled by methods that do not alter the randomization ratio rapidly.

Bashir et al. have implemented a two-block design where patients are randomized 1:1 in a group of 10 and then based on the outcomes, they randomized the next ten patient to the superior treatment \citep{bashir2012randomized}. 
However, at the second block, they randomized patients to the lower dosage because the probability of randomizing patient to the lower dosage was 0.9 compared to 0.1 to the higher dosage. 
This design should be considered a randomized control trial for the first block and an observational study for the second block. 

It is shown that using a small number of blocks (K = 2, 4 and 5) has a good tradeoff between efficiency and ethically treating patients to the best known superior treatment. 
A large number of blocks should be clearly avoided for reasons of both ethics and efficiency.
Traditional RAR does not only delay the trial but also affects the clinical conclusion achieved.
We have not considered the multiple treatment design with more than 2-arms.
The design would be much more complex and it should be examined further.

An R package (blockRAR), for the frequentist and Bayesian models, is implemented in R and released as open source software under the MIT license. 
The blockRAR package is available at Comprehensive R Archive Network (CRAN) and at \href{https://thevaachandereng.github.io/blockRAR/}{https://thevaachandereng.github.io/blockRAR/}. 
We used blockRAR version 1.0.0 for all analyses.

%\section{Support for \textsf{\journalclass}}
%We offer on-line support to participating authors. Please contact
%us via e-mail at \dots
%
%We would welcome any feedback, positive or otherwise, on your
%experiences of using \textsf{\journalclass}.

\begin{sm}
Supplementary material are available.
\end{sm}

\begin{dci}
None declared.
\end{dci}

\begin{funding}
None declared.
\end{funding}

\begin{acks}
We are grateful to Tom Cook for his helpful comments and suggestions.
\end{acks}

\bibliographystyle{SageV}
\bibliography{refs}

\end{document}